**Realization of the structural fluctuation of biomolecules in solution: Generalized Langevin Mode Analysis.**

Masatake Sugita[1,2] and Fumio Hirata[3*]

[1] Department of Bioinformatics, College of Life Sciences, Ritsumeikan University, 1-1-1, Noji-higashi, Kusatsu, Shiga, 525-8577, Japan
[2] Department of Computer Science, School of Computing, Tokyo Institute of Technology, W8-76, 2-12-1, Ookayama Meguro-ku, Tokyo, 152-8550, Japan
[3] Institute for Molecular Science, Okazaki 444-8585, Japan

**ABSTRACT**

A new theoretical method, referred to as Generalized Langevin Mode Analysis (GLMA), is proposed to analyze the mode of structural fluctuations of a biomolecule in solution. The method combines the two theories in the statistical mechanics, or the Generalized Langevin theory and the RISM/3D-RISM theory, to calculate the second derivative, or the Hessian matrix, of the free energy surface of a biomolecule in aqueous solution, which consists of the intramolecular interaction among atoms in the biomolecule and the solvation free energy.

The method is applied to calculate the wave-number spectrum of an alanine dipeptide in water for which the optical heterodyne-detected Raman-induced spectroscopy (RIKES) spectrum is available to compare with. The theoretical analysis reproduced the main features of the experimental spectrum with respect to the peak positions of the four bands around ~90 cm$^{-1}$, ~240 cm$^{-1}$, ~370 cm$^{-1}$, and 400 cm$^{-1}$, observed in the experimental spectrum, in spite that the physics involved in the two spectrum was not exactly the same: the experimental spectrum includes the contributions from the dipeptide and the water molecules interacting with the solute, while the theoretical one is just concerned with the solute molecule, influenced by solvation.

Two major discrepancies between the theoretical and experimental spectra, one in the band intensity around ~100 cm$^{-1}$, and the other in the peak positions around ~370 cm$^{-1}$, are discussed in terms of the fluctuation mode of water molecules interacting with the dipeptide, which is not taken *explicitly* into account in the theoretical analysis.

# INTRODUCTION

Function of protein is closely related to its structural fluctuation in the solution phase. [1-6] For instance, a ligand molecule cannot access to the binding pocket without fluctuation of the protein since the binding mode between the protein and its ligand is mostly very tight. It is also known that the change of the fluctuation originated from a ligand binding at the distant position from the active site affect to the enzymatic activity. [7-10] It indicates that the catalytic process of the enzyme is closely related to the inherent fluctuation of the protein. Since the fluctuation related to function of a protein is regarded as a long-period collective-mode, it is essential to elucidate the fluctuations with a long time scale at the atomic resolution for understanding the functional expression mechanism of the protein.

The dynamic behavior of proteins is experimentally analyzed by the spectroscopic methods such as NMR, [3,4] transient grating spectroscopy, [5,6] optical Kerr effect (OKE) spectroscopy. [11,12] These methods can reveal the characteristics of the fluctuation of the proteins. However, the resolution of the methods is not fine enough to understand the fluctuation in atomic detail. Therefore, it is a common maneuver to utilize a theoretical method such as molecular dynamics simulation (MD) in order to analyze fluctuations of the molecule in atomic resolution. [13-20]

The most typical and effective method for analyzing fluctuations of a biomolecule computationally is the quasi-harmonic analysis, or the principal component analysis, of the variance-covariance matrix of the trajectory obtained from the MD simulation. However, there are certain limitations even in the method based on MD. For instance, in order to analyze detailed aspects of fluctuations, sampling of the trajectory should be at least several times longer than the relaxation time of the fluctuation. However, a usual MD for proteins is able to sample several tens of microseconds under infinite dilution conditions. It is obvious that a MD simulation cannot catch such fluctuations in detail, as its time constant exceeding tens of microseconds. Furthermore, sampling of the configuration space becomes increasingly difficult when the solvent consists of several chemical components in addition to water, such as electrolytes and denaturants. It will be desired to accelerate the rate of sampling in the configurational space in one way or another.

Another popular method for analyzing the structural fluctuation of biomolecules is the Normal Mode Analysis (NMA). [21-23] （引用） The method gives the second derivative of the potential energy of a biomolecule, or Hessian, which represents the structural fluctuation of a molecule in terms of the frequency spectrum. The analysis provides the wide range of frequency spectrum of a molecule, ranging from the bond stretching-mode to that corresponding to the collective mode. Therefore, it is free from the concern about the length of trajectory of a MD simulation. However, this analysis has a fatal drawback. The analysis is performed in the environment of *vacuum*. So, the method in principle evaluates fluctuations in vacuum. The structural, fluctuation, and dynamics of a protein in solution phase is completely different from those in vacuum. The method will give an erroneous picture for the structural fluctuation of protein.

An insightful work which may solve the problem has been published by Kim and Hirata in 2013. [24] They have derived a generalized Langevin equation which represents the dynamics of a solute molecule conjugated with the density fluctuation of solvent. The most important point of the equation is that the shape of the equation is essentially a Langevin equation of a *harmonic oscillator* immersed in solvent or water. So, it has a general appearance similar to the Langevin equation treated by Wang and Uhlenbeck in 1945, and by Lamm and Szabo later in the context of Langevin mode analysis (LMA).[25,26] However, there are two essential differences between the new and old theories. The first of those is concerned with solvent in which the harmonic oscillator is immersed. The solvent in the old theories is a sort of structureless "gel", while that in the new theory is an assembly of molecules. The important terms of the equation, such as the friction and the random force, have a microscopic expression that can be realized with the aid of another generalized Langevin equation of fluid under a force exerted by the protein. Another difference from the old theories is in its restoring force proportional to the displacement of atoms from their equilibrium position. In the old theories, the proportional constant, or the Hessian, is the second derivative of the *potential energy* with respect to the atomic coordinate, while it is the second derivative of the free energy in the new theory, that includes the solvent induced force. The Hessian concerning the solvent induced force may be obtained from the RISM/3D-RISM theory as the second derivative of the solvation free energy. So, if one drops the memory and random force terms from the equation, it is formally identical to that of a harmonic oscillator employed in the normal mode analysis, but with the Hessian that includes solvent-induced force. The isomorphism between the two theories gives us a new possibility to analyze the structural fluctuation of a biomolecule in solution without being concerned with the notorious problem inherent to MD, or the sampling of configuration space.

Now, the problem of analyzing the structural fluctuation of a biomolecule in solution is reduced to a calculation of the Hessian including solvent-induced force. According to the Kim-Hirata theory, the Hessian can be obtained in principle from the RISM/3D-RISM theory as the second derivative of the free energy with respect to the atomic coordinates of the molecule. However, it is a non-trivial numerical problem to calculate the Hessian matrix, the number of elements of which amounts to a square of the number of atoms consisting the biomolecule: for a typical protein including ~$10^4$ atoms, the number of matrix elements becomes ~$10^8$.

In the present study, we propose a new method for evaluating the Hessian matrix based on the RISM/3D-RISM theory. In the method, we prepare a trajectory of a biomolecule in water by running a MD simulation in which each atom is driven by the interatomic forces as a derivative of the *free energy surface* with respect to the atomic coordinates.[27] The free energy surface is evaluated by means of the RISM/3D-RISM method. The Hessian matrix is obtained from the trajectory taking the second derivative of the free energy surface with respect to the atomic coordinate of the solute at each snapshot, and taking the average over the entire trajectory.

The method is applied to a rather small biomolecule, alanine dipeptide, in water to calculate the wave number spectrum of the molecule. The result is compared with the experimental data of the spectrum around the tera Hertz region observed by means of optical heterodyne-detected Raman-induced spectroscopy (RIKES).[11,12]

**Theory**

In the present study, we apply the generalized Langevin equation (GLE) of a protein in water at the infinite dilution, derived by Kim and Hirata, to calculate the structural fluctuation. The equation has been derived by projecting all the variables in the phase space, concerning both the protein and water, onto the four dynamic variables: the density field and its conjugated momentum of water, and the spatial and momentum coordinates of atoms of a single protein molecule. It should be noted that the canonical ensemble average for the projection is taken over the entire coordinates, spatial and momentum, of both the protein and water. The projection gives rise essentially to two equations, one for a protein molecule, and the other for the density field of water molecules, which are conjugated each other. In the present study, we are just concerned with the GLE for a protein, which reads

$$M_\alpha \frac{d^2 \Delta \mathbf{R}_\alpha(t)}{d^2 t} = -\sum_\beta A_{\alpha\beta} \cdot \Delta \mathbf{R}_\beta(t) - \int_0^t \sum_\beta \Gamma_{\alpha\beta}(t-s) \cdot \frac{\mathbf{P}_\alpha(s)}{M_\beta} + \mathbf{W}_\alpha(t) \quad (1)$$

where $\Delta \mathbf{R}_\alpha(t) = \mathbf{R}_\alpha(t) - \langle \mathbf{R}_\alpha \rangle$ denotes the displacement of an atom $\alpha$ of the protein from its equilibrium position $\langle \mathbf{R}_\alpha \rangle$. It is worthwhile to note that the ensemble average $\langle \cdots \rangle$ is taken over the entire phase space including both the protein and water. The second and third terms in the right-hand side in Eq. (1) are the friction and random forces acting on the atom $\alpha$ of the biomolecule as in the standard GLE, which play crucial roles in the dynamics or relaxation from a fluctuated structure to the equilibrium one. We ignore those two terms in this study in which we are just interested in the mode and magnitude of the fluctuation.

The first term in the right-hand-side is the restoring force acting on the atom $\alpha$ from the other atoms represented by $\beta$, which is proportional to the displacement of the atoms from their equilibrium position $\langle \mathbf{R}_\beta \rangle$, or the structural fluctuation. With those physics of GLE in mind, Eq. (1) is adapted specially to the analysis of structural fluctuation in the form,

$$M_\alpha \frac{d^2 \Delta \mathbf{R}_\alpha(t)}{d^2 t} = -\sum_\beta k_{\alpha\beta} \cdot \Delta \mathbf{R}_\beta(t) \quad (2)$$

where $k_{\alpha\beta}$ is the force-constant matrix, or the *Hessian* matrix, of the restoring force, and the following expression was derived by Kim and Hirata.

$$k_{\alpha\beta} = k_B T \langle \Delta \mathbf{R}_\alpha \Delta \mathbf{R}_\beta \rangle^{-1} \quad (3)$$

Eq. (2) is formally identical to the equation of a harmonic oscillator, so that we are able to use all the numerical recipe developed for the normal mode analysis (NMA) with a difference of physics involved in the two. Let us refer to the new method as "Generalized

Langevin Mode Analysis (GLMA)."

The equation (2) implies that the fluctuation of protein atoms is taking place on the energy surface which has a quadratic form with respect to the atomic coordinates, that is,

$$F(\Delta \mathbf{R}_1, \Delta \mathbf{R}_2, \cdots, \Delta \mathbf{R}_N) = \sum_\alpha \sum_\beta \Delta \mathbf{R}_\alpha A_{\alpha\beta} \Delta \mathbf{R}_\beta. \quad (4)$$

Kim and Hirata have made a physical *ansatz* that the energy should be the free energy consisting of the potential energy and the solvation free energy, not just a potential energy among the protein atoms, that is,

$$F(\Delta \mathbf{R}_1, \Delta \mathbf{R}_2, \cdots, \Delta \mathbf{R}_N) = U(\Delta \mathbf{R}_1, \Delta \mathbf{R}_2, \cdots, \Delta \mathbf{R}_N) + \Delta\mu(\Delta \mathbf{R}_1, \Delta \mathbf{R}_2, \cdots, \Delta \mathbf{R}_N) \quad (5)$$

where $U$ and $\Delta\mu$ are the potential energy and the solvation free energy. The ansatz is a logical consequence derived from Eq. (1) that concerns all the coordinates in the phase space including those of water molecules, not just those of protein. With the Eq. (3), the probability distribution of the displacement of atoms from the equilibrium state, or the structural fluctuation, is written as,

$$p(\Delta \mathbf{R}_1, \Delta \mathbf{R}_2, \cdots, \Delta \mathbf{R}_N) = \sqrt{\frac{|k|}{(2\pi)^{3N}}} \exp\left[-\frac{1}{2}\sum_\alpha \sum_\beta \Delta \mathbf{R}_\alpha k_{\alpha\beta} \Delta \mathbf{R}_\beta\right]. \quad (6)$$

The variance-covariace matrix is defined by

$$\langle \Delta \mathbf{R}_\alpha \Delta \mathbf{R}_\beta \rangle = \int_{-\infty}^{\infty} \cdots \int_{-\infty}^{\infty} \Delta \mathbf{R}_\alpha \Delta \mathbf{R}_\beta p(\Delta \mathbf{R}_1, \Delta \mathbf{R}_2, \cdots, \Delta \mathbf{R}_N) d\Delta \mathbf{R}_1 d\Delta \mathbf{R}_2 \cdots d\Delta \mathbf{R}_N. \quad (7)$$

The Hessian matrix can be obtained from Eq. (4) as the second derivative of the free energy surface,

$$k_{\alpha\beta} = \frac{\partial^2 F}{\partial \Delta \mathbf{R}_\alpha \partial \Delta \mathbf{R}_\beta} \quad (8)$$

So, with the definition of the Hessian matrix by Eq. (8), the mathematical isomorphism between NMA and GLMA was established. In the case of NMA, the free energy $F(\Delta \mathbf{R}_1, \Delta \mathbf{R}_2, \cdots, \Delta \mathbf{R}_N)$ should be replaced just by the potential energy $U(\Delta \mathbf{R}_1, \Delta \mathbf{R}_2, \cdots, \Delta \mathbf{R}_N)$.

The mathematical task of NMA or GLMA is to perform a principal-axis (component) analysis, or to solve an eigen value problem, of the Hessian matrix (Eq. (8)) or the variance-covariance matrix defined by Eq. (7), which are related through Eq. (3). However, there is a big stumbling block before we are able to perform the principal-axis analysis, that is how to realize the Hessian or variance-covariance matrix.

*Determining the Hessian matrix from the 3D-RISM/MD simulation*

From Eq. (5) and (8), the Hessian matrix of the free energy surface is split into two terms,

$$k_{\alpha\beta} = \frac{\partial^2 F}{\partial \Delta \mathbf{R}_\alpha \partial \Delta \mathbf{R}_\beta} \\ = \frac{\partial^2 U}{\partial \Delta \mathbf{R}_\alpha \partial \Delta \mathbf{R}_\beta} + \frac{\partial^2 \Delta \mu}{\partial \Delta \mathbf{R}_\alpha \partial \Delta \mathbf{R}_\beta}. \qquad (9)$$

The first term is the second derivative of the potential energy among atoms in the biomolecule, the calculation of which can be performed with a routine implemented in a standard program of MD simulation, such as AMBER. However, it is not a trivial problem to calculate the second term, because it is concerned with the solvation free energy of the biomolecule in solution. If one tries to do it with the MD simulation, it will be a disaster. Even to determine the solvation free energy of a single conformation of a biomolecule by a MD simulation will require a huge amount of computation time. However, it is relatively an easy task for the RISM/3D-RISM formalism thanks to a tactics proposed by Yu and Karplus, which enables to determine the derivatives of the correlation functions along the iteration process for solving the integral equations.[28]

The procedure begins with the 3D-RISM formula for the force acting on an atom of protein from the solvent molecules, derived by Yoshida and Hirata, which requires the derivative of the solvation free energy with respect to protein atoms.[27,29]

$$\frac{\partial \Delta \mu}{\partial \mathbf{R}_\alpha} = \sum_j \rho_j \int \left( \frac{\partial u_j^{uv}}{\partial \mathbf{R}_\alpha} \right) g_j^{uv}(\mathbf{r}) d\mathbf{r}, \qquad (10)$$

where $\rho_j$ denotes the density of the *j*-th atom in the solvent molecule, $u_j^{uv}(\mathbf{r})$ the interaction energy between the solute atom $\alpha$ and the solvent atom *j*, $g_j^{uv}(\mathbf{r})$ the spatial distribution of the solvent atom *j* around the protein. It should be noted that the both are an implicit function of the coordinates $\mathbf{R}_\alpha$ of the solute. The relation can be understood as a classical analogue of the Hellmann-Feynman theorem in the quantum mechanics if one regards $\mathbf{R}_\alpha$, $u_j^{uv}$, and $g_j^{uv}$ as the coordinate of a nucleus, the electronic energy, and the probability distribution of the electron which is a square of the electronic wave-function.

The second derivative of $\Delta\mu$ with respect to the protein coordinate is derived simply from Eq. (10) as

$$\frac{\partial^2 \Delta\mu}{\partial \mathbf{R}_\alpha \partial \mathbf{R}_\beta} = \sum_j \rho_j \int \frac{\partial^2 u_j^{uv}(\mathbf{r})}{\partial \mathbf{R}_\alpha \partial \mathbf{R}_\beta} g_j^{uv}(\mathbf{r}) d\mathbf{r} + \sum_j \rho_j \int \frac{\partial u_j^{uv}(\mathbf{r})}{\partial \mathbf{R}_\alpha} \frac{\partial g_j^{uv}(\mathbf{r})}{\partial \mathbf{R}_\beta} d\mathbf{r} \quad (11)$$

Eq. (11) includes the derivative of the spatial distribution function which makes the calculation potentially non-trivial. However, it can be performed by the Yu-Karplus method[28] along the course of iteration for solving the RISM/3D-RISM equation as follows.

Although the method is common to any closure to solve the equation, here, we just present the procedure corresponding to the Kovalenko-Hirata closure. Then, the RISM/3D-RISM equation consists of the two equations, which are written as

$$h_j^{uv}(\mathbf{r}) = \sum \int \chi_{jl}^{vv}(|\mathbf{r}-\mathbf{r}'|) c_l^{uv}(\mathbf{r}') d\mathbf{r}'$$
$$\equiv \chi_{jl}^{vv} * c_l^{uv}(\mathbf{r}) \quad (12)$$

$$h_j^{uv}(\mathbf{r}) = \begin{cases} \exp[d_j^{uv}(\mathbf{r})] - 1 & \text{for } d_j^{uv}(\mathbf{r}) \leq 0 \\ d_j^{uv}(\mathbf{r}) & \text{for } d_j^{uv}(\mathbf{r}) > 0 \end{cases} \quad (13)$$

$$d_j^{uv}(\mathbf{r}) \equiv -u_j^{uv}(\mathbf{r})/k_B T + h_j^{uv}(\mathbf{r}) - c_j^{uv}(\mathbf{r}) \quad (14)$$

where $h_j^{uv}(\mathbf{r}) \equiv g_j^{uv}(\mathbf{r}) - 1$. If one interprets the equation in terms of the non-linear response theory, $\chi_{lj}^{vv}$ is the site-site pair correlation function of solvent that plays the susceptibility or response function to the perturbation $c_j^{uv}(\mathbf{r})$ from the solute molecule.[30] The derivative of the correlation functions with respect to the atomic coordinate of protein can be written as

$$\frac{\partial h_j^{uv}(\mathbf{r})}{\partial \mathbf{R}_\alpha} = \sum_j \chi_{jl}^{vv} * \frac{\partial c_j^{uv}}{\partial \mathbf{R}_\alpha} \tag{15}$$

$$\frac{\partial h_j^{uv}(\mathbf{r})}{\partial \mathbf{R}_\alpha} = \begin{cases} \left(-\frac{1}{k_B T}\frac{\partial u_j^{uv}(\mathbf{r})}{\partial \mathbf{R}_\alpha} + \frac{\partial t_j^{uv}(\mathbf{r})}{\partial \mathbf{R}_\alpha}\right)\exp\left(-u_j^{uv}(\mathbf{r})/k_B T + t_j^{uv}(\mathbf{r})\right) \\ \qquad\qquad\qquad\qquad\qquad \text{for} \quad -u_j^{uv}(\mathbf{r})/k_B T + t_j^{uv}(\mathbf{r}) \leq 0 \\ -\frac{1}{k_B T}\frac{\partial u_j^{uv}(\mathbf{r})}{\partial \mathbf{R}_\alpha} + \frac{\partial t_j^{uv}(\mathbf{r})}{\partial \mathbf{R}_\alpha} \quad \text{for} \quad -u_j^{uv}(\mathbf{r})/k_B T + t_j^{uv}(\mathbf{r}) > 0 \end{cases}$$

$$\tag{16}$$

The derivatives can be calculated along the course of iteration to find the solutions for the correlation functions themselves.

In order to compare the theoretical results with experimental data of the wavenumber spectrum, the Hessian matrix is diagonalized to find the eigen value and vector for each snap of the simulation, and averaged over the trajectory to take account for the thermal fluctuation of solute conformation.

**Computational details**

To calculate the Hessian matrix, $k_{\alpha\beta}$, we first perform the molecular dynamics simulations based on the MTS-MD/OIN/GSFE/3D-RISM-KH (MD/3D-RISM-KH) program [31] implemented in the AMBER 18 software package.

In the MD/3D-RISM-KH simulation, we use the reference system propagator algorithm (RESPA) proposed by Omelyan and Kovalenko, in which the calculation of the solvent induced force is performed at every *rismnrespa* steps, not at every step, of the MD simulation in order to save the computation time.[32] So, the solvent induced force (Eq. (10)) is calcuated in every *rismnrespa* steps. Furthermore, the mean force induced by the solvent is extrapolated in *fcestride* steps. Thus, 3D-RISM calculations are performed every *fcestride* × *rismnrespa* steps.

The ff14sb force field parameter is selected. Water molecules are represented by the modified SPC/E model. The cutoff radius of the solute-solvent interactions is set to 14 Å. The 3D-RISM-KH integral equations were discretized on a rectangular grid of resolution $\delta r = 0.5$ Å and converged to a relative root-mean-square residual tolerance of $\delta\varepsilon = 10^{-4}$ by using the MDIIS accelerated numerical solver.

For the simulation of the alanine dipeptide, *fcestride*, *fcenbas,* and *fcenbasis* is set to 10, 20, and 200, respectively. An 80 ns simulation was performed using these parameters. For the simulation of the Met-enkephalin, *fcestride*, *fcenbas,* and *fcenbasis* is set to 10, 40, and 500, respectively. A 200 ns simulation was performed using these parameters.

**Results and Discussions**

***3D-RISM/MD trajectory of the dipeptide projected onto dihedral-angle space:*** Depicted in Fig. 1 are distributions of the MD trajectory over 200 ns projected onto the angular coordinate space spanned by the two dihedral angles ($\psi_1, \varphi_2$) illustrated in the figure. A measure peak of the distribution is found around $\psi_1 = 150°$ and $\varphi_2 = -60°$, marked by (ii) in Fig. 1 (b), which corresponds roughly to the *trans-gauche* conformation. There is also a minor peak of the distribution around $\psi_1 = 150°$ and $\varphi_2 = -150°$, which corresponds roughly to the *trans-trans* conformation.

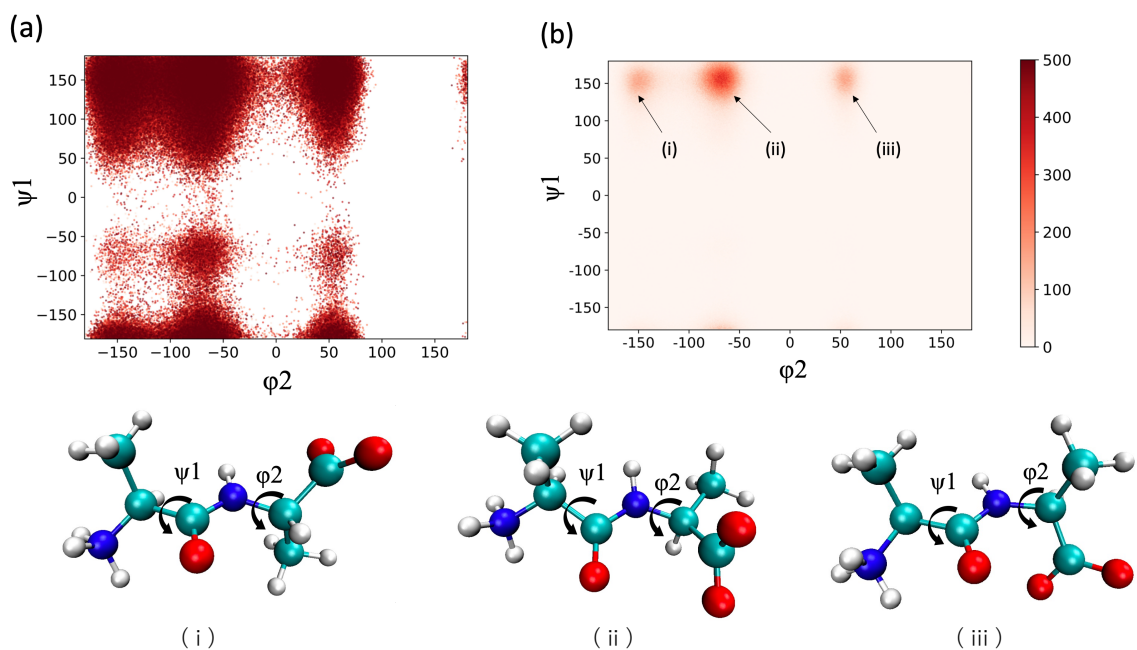

**Fig. 1**. **The trajectory projected onto the dihedral angle ($\Psi 1, \varphi 2$) space: (a) the snapshots at every 80 steps, (b) the distribution of the trajectory. The molecular pictures depicted under the two figures (a) and (b) illustrate the conformations corresponding to the snapshots**

**(i), (ii), and (iii) in Fig. (b).**

*Spectrum from multiple snapshots:* Shown in Fig. 2 is the wavenumber spectrum of an alanine-dipeptide in water, obtained from the Hessian matrix, $k_{\alpha\beta}$, by diagonalizing the matrix and averaging over 1000 snapshots. The snapshots are taken from an 80ns 3D-RISM/MD trajectory, evenly spaced. The structure of each snapshot was not minimized for the free energy.

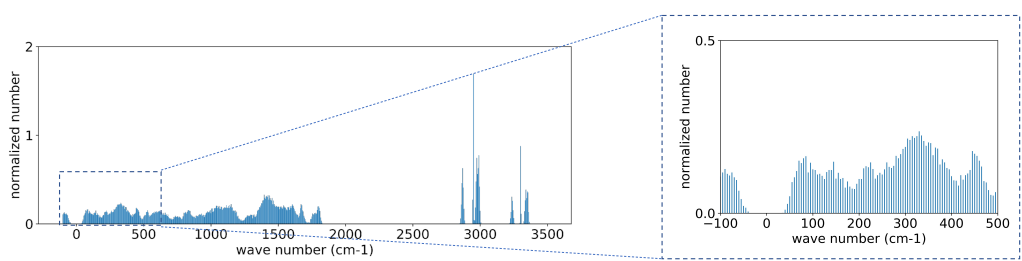

**Fig. 2 The wavenumber spectrum of an alanine-dipeptide in water. The inset is exploded view of the spectrum in the range of wave number from -100 cm$^{-1}$ to 500 cm$^{-1}$.**

The spectrum can be discussed in the four different regions: (1) the region greater than ~2700 cm$^{-1}$, (2) the region between ~500 to ~1800 cm$^{-1}$, (3) the region less than ~500 cm$^{-1}$, (4) the region of negative wave number. The regions greater than ~2700 cm$^{-1}$ is attributed apparently to stretching and bending of chemical-bonds related to hydrogen atoms. On the other hand, the spectrum between ~500 to ~1800 cm$^{-1}$ seems to be assigned to the local oscillations, stretching and bending, with which the heavy atoms such as C, O, N are concerned. Those vibrational modes are largely determined by the model of interatomic interaction-potential employed by the simulation, and they are not of much interest for the structural fluctuation of biomolecules. It is the regions (3) and (4) that is closely related to the structural fluctuation. The exploded view of the spectrum in that wave-number region is depicted in the right figure. So, we focus our attention on that region of the spectrum in the following.

Depicted in Fig. 3 is the spectrum of the dipeptide in the wave-number less than 500 cm$^{-1}$, which is compared with a corresponding experimental data obtained by Klaas and his coworkers by means of optical heterodyne-detected Raman-induced spectroscopy

(RIKES).[11,12] Although there are apparent differences observed between the two results, there is a common feature in the two spectra. The two spectra have four peaks between the wave numbers, 0 to 500 cm$^{-1}$, which are relatively close each other, that is, ~ 90 cm$^{-1}$, 250 cm$^{-1}$, 370 cm$^{-1}$ and 450 cm$^{-1}$ in the RIKES spectra, while ~90 cm$^{-1}$, ~240 cm$^{-1}$, ~320 cm$^{-1}$, and ~450 cm$^{-1}$ in the GLMA spectra.

There are marked differences between the two spectra in the following respects. (1) the negative frequency region observed in the GLMA result, which is absent in the experimental data; (2) small subpeak around 0 cm$^{-1}$ seen in the RIKES result, which is absent in the GLMA spectra; (3) the large difference in the intensity between the two spectra, especially around 0 to 100 cm$^{-1}$ region; (4) relatively large difference, about 50 cm$^{-1}$, in the peak positions around 320 cm$^{-1}$.

The spectrum at the negative frequency region in the theoretical result is apparently reflecting the representative points in the $\phi-\psi$ plot, which are assigned to the transient regions among the three stable conformations of the dipeptide molecule, depicted in Fig. 1. The points represent rare events, and there is no way to be observed by the spectrum in the real world. On the other hand, the theoretical analysis can observe such events, since the points reflect merely the negative-curvature region of the quadratic surface, over which the system pass over quickly. The analysis of such a region of the "transition state" is of great interest from a view point of the kinetic analysis of a chemical reaction. (In the present case, the chemical reaction is an "isomerization") Nevertheless, we just focus our attention on the structural fluctuation of the biomolecule, projected on the *real* frequency in the present study.

The second to fourth differences are mainly caused by the discrepancy in the physics involved in the two analyses. According to the authors of the experimental paper, the spectral data was obtained by subtracting the intensity concerning pure solvent from overall spectrum including solute and solvent.[11,12] Therefore, the intensity depicted in the figure includes two contributions, one from the solute and the other from water molecules that are interacting with the solute. The small subpeak seen around 0 cm$^{-1}$ is likely to be assigned to the *diffusive motion* of water molecules interacting with the dipeptide. On the other hand, the theoretical analysis does not include any spectrum contributed from water, although the contribution from solute-solvent interaction is included implicitly in the spectrum of the solute. It may be the reason why the theoretical spectrum does not have the intensity around 0 cm$^{-1}$. The experimental spectra may also be contributed from the translational and rotational diffusion of the dipeptide. Such degrees of freedom are also removed from the theoretical analysis.

The same reason is attributed to the large difference in the intensity between the two spectra, especially those around ~100 cm$^{-1}$ region. The region of the RIKES spectrum includes contributions from water molecules interacting with the solute. Those water molecules interacting with the solute, especially via hydrogen-bond, are likely to be involved in oscillatory motions in lower frequency modes. So, it is suggested that the large intensity around 100 cm$^{-1}$ is assigned to the intermolecular oscillatory motion of water molecules interacting with the solute.

The rather large discrepancy in the peak positions between the theoretical and the experimental spectra, ~ 320 cm$^{-1}$ vs. ~ 370 cm$^{-1}$, may require a structural analysis of the mode of fluctuation.

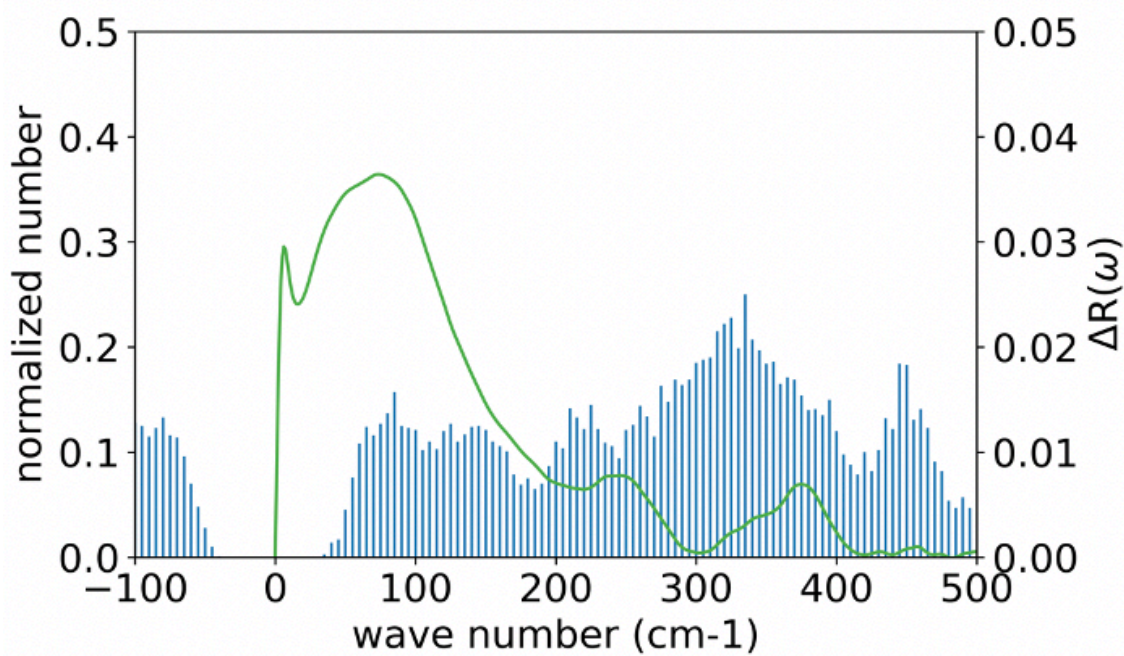

**Fig. 3 Comparing the low frequency spectrum from GLMA with that from RIKES**

***Contributions to the spectra from different conformations.*** Illustrated in Fig. 4 is the structure and the fluctuational mode of the dipeptide corresponding to the peak positions in the GLMA spectrum, which are obtained by diagonalizing the Hessian matrix. In the figures, the direction and amplitude of the fluctuation of each atom are illustrated by thick arrows.

As can be seen, all the modes carry a collective character, more or less, in the sense that the oscillations extend over the entire molecule. For example, the mode with the lowest frequency, or 87 cm$^{-1}$, looks like a "hinge-bending" motion around the C=O carbonyl bond, since the central carbonyl group and the two terminal groups are oscillating in the opposite phases. On the other hand, the mode at 452 cm$^{-1}$ seems to be more localized around the N-terminus group. It may be the reason why the frequency is relatively high.

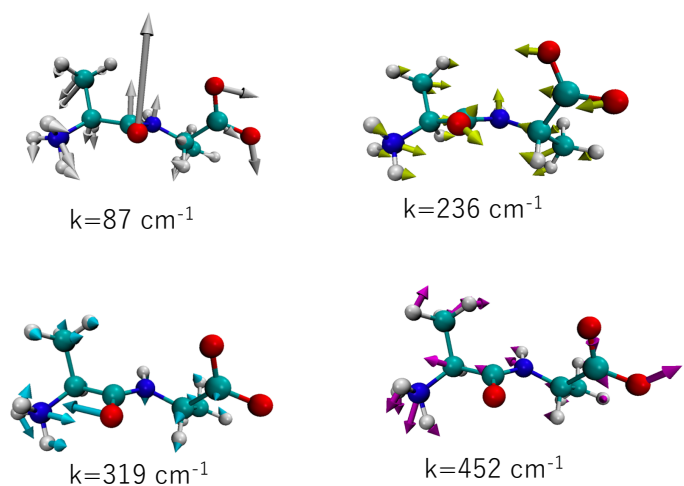

**Fig. 4 The modes of fluctuation corresponding to the peaks of wave number spectrum.**

An interesting behavior is seen in the mode assigned to 319 cm$^{-1}$, in which the carbonyl and the amide nitrogen are oscillating in the opposite phase. The oscillation is indicative of a water molecule bridging between the two atoms through a hydrogen-bond. In order to clarify if it is the case or not, the solvation structure of the molecule in that fluctuation mode, or k=319 cm$^{-1}$, was analyzed.

Depicted in Fig. 5 are the radial distribution functions (RDF) of water molecules around the carbonyl oxygen and the amide nitrogen of the dipeptide. The sharp peak at $r$~1.8 A in the O(peptide)-H(water) RDF in Fig. 5 (c) is a manifestation of the hydrogen-bond between the carbonyl oxygen and the water hydrogen. The sharp peak at $r$~2.8 A in the N(peptide)-O(water) RDF in Fig. 5(a) is indicative of a strong electrostatic-interaction between the amide nitrogen and the water oxygen. The two sharp peaks in RDFs are strong evidence of the existence of the water-bridge between the carbonyl-oxygen and the amide-nitrogen through those strong interactions. The situation is illustrated by the cartoon at bottom right in Fig. 5. The analysis suggests that the water-bridge through the two strong interactions may be the origin of the fluctuation mode assigned to 319 cm$^{-1}$.

Now, we are at the position to answer the question raised at the end of the last subsection, namely, the origin of the difference in the peak positions at ~319 cm$^{-1}$ and ~370 cm$^{-1}$, respectively, in the theoretical and experimental spectra. As we have already mentioned, the experimental spectrum includes both the contributions from the dipeptide and water molecules, while the theoretical one is concerned only with the interactions within the dipeptide, which include contributions from solvent *implicitly*. Therefore, the peak in the theoretical spectrum at ~319 cm$^{-1}$ is originated mainly by the H(peptide)-O(peptide) interaction bridged by a water molecule. On the other hand, the peak in the experimental spectrum at ~370 is a composite band consisting of the contributions from the H(peptide)-O(peptide) interaction and the water molecule bridging the two atoms in the peptide. The frequency of the mode of the water molecule may be higher than that of H(peptide)-O(peptide) interaction, because the water molecule is connected with the two atoms in the peptide through the two strong interactions.

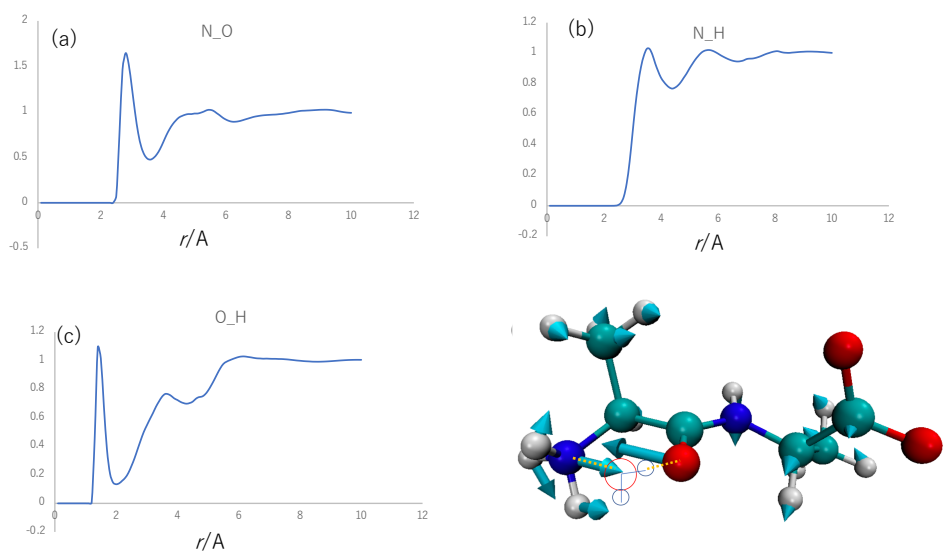

**Fig. 5 The radial distribution functions (RDF) of water molecules around the dipeptide in the fluctuational mode with 319 cm$^{-1}$. The picture in the bottom-right illustrates the water-bridge through the hydrogen-bond between O(peptide) and H(water), and the electrostatic interaction between N(peptide) and O(water).**

**Concluding remark**

A new theoretical method, referred to as Generalized Langevin Mode Analysis (GLMA), was proposed to analyze the mode of structural fluctuations of a biomolecule in solution. The method combines the two theories in the statistical mechanics, or the Generalized Langevin theory and the RISM/3D-RISM theory, to calculate the second derivative, or the Hessian matrix, of the free energy surface of a biomolecule in solution, which consists of the intramolecular interaction among atoms in the biomolecule and the solvation free energy.

The method was applied to the alanine dipeptide in aqueous solution to calculate the wave number spectrum of the system, especially focusing on the collective modes of the fluctuation less than 500 cm$^{-1}$, for which an experimental data by means of optical heterodyne-detected Raman-induced spectroscopy (RIKES) is available. The theoretical analysis reproduced the main features of the experimental spectrum with respect to the peak positions of the four bands around ~90 cm$^{-1}$, ~240 cm$^{-1}$, ~370 cm$^{-1}$, and 400 cm$^{-1}$, observed in the experimental spectrum, in spite that the physics involved in the two spectrum was not exactly the same: the experimental spectrum includes the contributions from the dipeptide and the water molecules interacting with the solute, while the theoretical one is just concerned with the solute molecule.

There are two major discrepancies between the two spectra: one in the band intensity around ~100 cm$^{-1}$, and the other in the peak positions, that is, ~320 cm$^{-1}$ in the theory, while ~370 cm$^{-1}$ in the experiment. The large band intensity around ~100 cm$^{-1}$ in the RIKES spectrum was attributed to the oscillatory motion of water molecules interacting with the dipeptide, which is not included in the spectrum from GLMA. In order to examine the shift of the peak positions from ~320 cm$^{-1}$ in the GMLA spectrum to ~370 cm$^{-1}$ in the RIKES spectrum, the radial distribution function of water around the dipeptide, which is in the fluctuation mode around ~320 cm$^{-1}$, was analyzed. From the analysis, it was concluded that the shift of the peak position from ~320 cm$^{-1}$ in the GLMA spectrum to ~370 cm$^{-1}$ in the RIKES spectrum is attributed to the oscillatory motion of a water molecule which bridges between two atoms, N and O, in the dipeptide, through the hydrogen-bond between H(water) and O(peptide), and the electrostatic interaction between N(peptide) and O(water).

The present study is indicative that the fluctuational mode of water molecules interacting with the peptide should be take into consideration in order for getting better agreement with the RIKES data. Such a goal may be achieved by solving the generalized

Langevin equation for water molecules interacting with the solute molecule. Such an equation has been already derived by Kim and Hirata, along with the generalized Langevin equation for a solute in water, which was employed in the present paper to analyze the structural fluctuation of the dipeptide.

The GLMA method developed here can be applied to explore the structural fluctuation of protein in solution without any further development in the theory, but with an assist of much greater computational power. It is our future plan to carry out such studies concerning the structural fluctuation of protein.

**Acknowledgement**

The authors gratefully acknowledge Klaas Wynne for providing the RIKES data. They are also grateful to Hideaki Shirota for bringing Wynne's paper to our attention. This work was partly supported by JSPS KAKENHI Grant Number 30737523. The part of the work was carried out while FH was a fellow of Toyota Physical and Chemical Research Institute (Toyota Riken).

(1) Gekko, K.; Obu, N.; Li, J.; Lee, J. C. A Linear Correlation Between the Energetics of Allosteric Communication and Protein Flexibility in the. *Biochemistry* **2004**, *43*, 3844–3852.

(2) Gekko, K.; Kamiyama, T.; Ohmae, E.; Katayanagi, K. Single Amino Acid Substitutions in Flexible Loops Can Induce Large Compressibility Changes in Dihydrofolate Reductase. *J. Biochem* **2000**, *128*, 21–27.

(3) Bhabha, G.; Lee, J.; Ekiert, D. C.; Gam, J.; Wilson, I. A.; Dyson, H. J.; Benkovic, S. J.; Wright, P. E. A Dynamic Knockout Reveals That Conformational Fluctuations Influence the Chemical Step of Enzyme Catalysis. *Science* **2011**, *332*, 234–238.

(4) Akasaka, K. Probing Conformational Fluctuation of Proteins by Pressure

Perturbation. *Chem. Rev.* **2006**, *106,* 1814–1835.

(5) Kuroi, K.; Okajima, K.; Ikeuchi, M.; Tokutomi, S.; Terazima, M. Transient Conformational Fluctuation of TePixD During a Reaction. *Proceedings of the National Academy of Sciences* **2014**, *111* (41), 14764–14769.

(6) Tanaka, K.; Nakasone, Y.; Okajima, K.; Ikeuchi, M.; Tokutomi, S.; Terazima, M. Oligomeric-State-Dependent Conformational Change of the BLUF Protein TePixD (Tll0078). *Journal of Molecular Biology* **2009**, *386* (5), 1290–1300.

(7) Monod, J.; Wyman, J.; Changeux, J.-P. On the Nature of Allosteric Transitions: a Plausible Model. *Journal of Molecular Biology* **1965**, *12* (1), 88–118.

(8) D E Koshland, J.; Némethy, G.; Filmer, D. Comparison of Experimental Binding Data and Theoretical Models in Proteins Containing Subunits*. *Biochemistry* **1966**, *5* (1), 365–385.

(9) Srinvasan, B.; Forouhar, F.; Shukla, A.; Sampangi, C.; Kulkarni, S.; Abashidze, M.; Seetharaman, J.; Lew, S.; Mao, L.; Acton, T. B.; et al. Allosteric Regulation and Substrate Activation in Cytosolic Nucleotidase II From Legionella Pneumophila. *FEBS J.* **2014**, *281*, 1613–1628.

(10) Cuendet, M. A.; Weinstein, H.; LeVine, M. V. The Allostery Landscape: Quantifying Thermodynamic Couplings in Biomolecular Systems. *J. Chem. Theory Comput.* **2016**, *12*, 5758–5767.

(11) Giraud, G.; Wynne, K. Time-Resolved Optical Kerr-Effect Spectroscopy of Low-Frequency Dynamics in Di- L-Alanine, Poly- L-Alanine, and Lysozyme in Solution. *J. Am. Chem. Soc.* **2002**, *124* (41), 12110–12111.

(12) Giraud, G.; Karolin, J.; Wynne, K. Low-Frequency Modes of Peptides and Globular Proteins in Solution Observed by Ultrafast OHD-RIKES Spectroscopy. *Biophys J.* **2003**, 1–11.

(13) Kitao, A.; Hirata, F.; Go, N.; The effect of solvent on the conformation and the collective motions of protein: normal mode analysis and molecular dynamics simulations of melittin in water and in vacuum. Chem. Phys. **1991**.*158*. 447-472.

(14) Kitao, A.; Hayward, S.; Go, N. Energy Landscape of a Native Protein: Jumping-Among-Minima Model. *Proteins* **1998**, *33*, 496–517.

(15) Kitao, A.; Go, N. Investigating Protein Dynamics in Collective Coordinate Space. **1999**, 1–6.

(16) Karasawa, N.; Mitsutake, A.; Takano, H. Identification of Slow Relaxation

Modes in a Protein Trimer via Positive Definite Relaxation Mode Analysis. *J. Chem. Phys.* **2019**, *150*, 084113.

(17) Garcia, A. E. Large-Amplitude Nonlinear Motions in Proteins. *Phys. Rev. Lett.* **2011**, *68* (17), 2696–2700.

(18) Schwantes, C. R.; Pande, V. S. Improvements in Markov State Model Construction Reveal Many Non-Native Interactions in the Folding of NTL9. *J. Chem. Theory Comput.* **2013**, *9*, 2000–2009.

(19) Mori, T.; Saito, S. Dynamic Heterogeneity in the Folding/ Unfolding Transitions of FiP35. *J. Chem. Phys.* **2015**, *142*, 135101.

(20) Pérez-Hernández, G.; Paul, F.; Giorgino, T.; De Fabritiis, G.; Noé, F. Identification of Slow Molecular Order Parameters for Markov Model Construction. *J. Chem. Phys.* **2013**, *139*, 015102.

(21) Sakuraba, S.; Joti, Y.; Kitao, A. Detecting Coupled Collective Motions in Protein by Independent Subspace Analysis. *J. Chem. Phys.* **2010**, *133*, 185102.

(22) Harmonic Dynamics of Proteins: Normal Modes and Fluctuations in Bovinepancreatictrypsin. *Proceedings of the National Academy of Sciences* **1983**, *80*, 6571–6575.

(23) Go, N.; Nishikawa, T. Dynamics of a Small Globular Protein in Terms of Low-Frequency Vibrational Modes. *Proceedings of the National Academy of Sciences* **1983**, *80*, 3696–3700.

(24) Kim, B.; Hirata, F. Structural Fluctuation of Protein in Water Around Its Native State: a New Statistical Mechanics Formulation. *J. Chem. Phys.* **2013**, *138* (5), 054108–054112.

(25) Wang M. C.; Uhlenbeck. On the theory of Brownian motion II. *Rev. Mod. Phys.* **1945**, *17*, 323-342.

(26) Lamm, G.; Szabo, A. Langevin mode of macromolecules. *J. Chem. Phys.* **1986**, *85*, 7334-????.

(27) Miyata, T.; Hirata, F. Combination of Molecular Dynemiacs Method and 3D-RISM Theory for Conformational Sampling of Large Flexible Molecules in Solution. *J. Comput. Chem.* **2007**, *29*, 871-882.

(28) Yu, H.-A.; Karplus, M. J. Chem. Phys., 1988, 89, 2366-????.

(29) Yoshida, N.; Hirata, F. A New Method to Determine Electrostatic Potential Around a Macromolecule in Solution from Molecular Wave Functions. *J.*

*Comput. Chem.* **2006**, *27* (4), 453-462.

(30) *Exploring Life Phenomena with Statistical Mechanics of Molecular Liquids.* Hirata, F., CRC press, 2021.

(31) Omelyan, I.; Kovalenko, A. MTS-MD of Biomolecules Steered with 3D-RISM-KH Mean Solvation Forces Accelerated with Generalized Solvation Force Extrapolation. *J. Chem. Theory Comput.* **2015**, *11* (4), 1875–1895.